\documentclass[12pt]{iopart}
\usepackage{graphicx} 
\usepackage{float} 
\usepackage{amssymb, bm} 
\usepackage{xcolor}
\begin{document}

\title{Generative adversarial networks for data-scarce spectral applications}

\author{J.~J. Garc\'{\i}a-Esteban}
\address{Departamento de F\'{\i}sica Te\'orica de la Materia Condensada and Condensed Matter Physics Center (IFIMAC),
Universidad Aut\'onoma de Madrid, E-28049 Madrid, Spain}
\author{J.~C. Cuevas}
\address{Departamento de F\'{\i}sica Te\'orica de la Materia Condensada and Condensed Matter Physics Center (IFIMAC),
Universidad Aut\'onoma de Madrid, E-28049 Madrid, Spain}
\author{J. Bravo-Abad}
\address{Departamento de F\'{\i}sica Te\'orica de la Materia Condensada and Condensed Matter Physics Center (IFIMAC),
Universidad Aut\'onoma de Madrid, E-28049 Madrid, Spain}

\date{\today}

\begin{abstract}
Generative adversarial networks (GANs) are one of the most robust and versatile techniques in the field of generative
artificial intelligence.~In this work, we report on an application of GANs in the domain of synthetic spectral data
generation, offering a solution to the scarcity of data found in various scientific contexts. We demonstrate the proposed
approach by applying it to an illustrative problem within the realm of near-field radiative heat transfer involving a
multilayered hyperbolic metamaterial. We find that a successful generation of spectral data requires two modifications 
to conventional GANs:~\emph{(i)} the introduction of Wasserstein GANs (WGANs) to avoid mode collapse, and, \emph{(ii)} 
the conditioning of WGANs to obtain accurate labels for the generated data. We show that a simple feed-forward neural network (FFNN), when augmented 
with data generated by a CWGAN, enhances significantly its performance under conditions of limited data availability, 
demonstrating the intrinsic value of CWGAN data augmentation beyond simply providing larger datasets. In addition, we
show that CWGANs can act as a surrogate model with improved performance in the low-data regime with respect 
to simple FFNNs.~Overall, this work highlights the potential of generative machine learning algorithms in scientific
applications beyond image generation and optimization.
\end{abstract}

\maketitle

\section{Introduction} \label{sec-intro}
Machine learning, a rapidly expanding area within computer science, focuses on advancing the foundations and technology 
that enables machines to learn from data~\cite{Mitchell1997,LeCunDeepReview,Goodfellow2016Book,Aggarwal2018}. Deep 
learning, on the other hand, is a subset of machine learning techniques that uses artificial neural networks (ANNs) to 
model and solve complex data-driven problems, such as image and speech recognition~\cite{Krizhevsky2012,Hinton2012Oct}, natural 
language processing~\cite{Cho2014Jun}, and autonomous driving~\cite{Shalev-Shwartz2016Oct}, among  many others. With the current explosion of data and the rapid development of improved hardware and
algorithms, machine learning and deep learning are becoming crucial tools in many industries, including healthcare, 
finance, and manufacturing.

Motivated by this success, machine learning and deep learning techniques are attracting increasing attention from a 
variety of scientific disciplines beyond computer science, revolutionizing traditional approaches to the modeling and
analysis of data-driven scientific problems.~In physics, these techniques are employed to tackle complex problems~\cite{Carleo2019}, 
including the representation of quantum many-body wave functions~\cite{Carleo2017}, the discovery and identification of phase transitions in condensed-matter systems~\cite{Wang2016,Carrasquilla2017,vanNieuwenburg2017}, the solution of  statistical problems~\cite{Wu2019}, the development of novel quantum information technologies~\cite{Dunjko2018}, the modeling of gravitational waves~\cite{Green2020}, and 
the design of nanophotonic devices with novel or improved functionalities~\cite{Peurifoy2018,So2020,Jiang2021,Frising2023}. Machine learning algorithms 
have been effectively used for the accelerated discovery and design of new materials and molecules in the fields of materials science and
chemistry~\cite{Butler2018,GomezBombarelli2018}, being instrumental in molecular dynamics simulations~\cite{Zhang2018}, in predicting chemical reactions~\cite{Coley2017}, and in modeling the quantum mechanical energies of molecules~\cite{Rupp2012}.~In the realm 
of biology, the applications of machine learning and deep learning are also vast, including breakthroughs in gene expression prediction tasks, prediction of micro-RNAs targets, and novel single-cell methods~\cite{Ching2018}. Overall, despite significant challenges, such as the interpretability and 
transferability, machine learning and deep learning are playing an increasingly pivotal role in the advance of a 
broad variety of scientific research methodologies.

In this context, generative adversarial networks (GANs), a powerful subset of generative machine learning, have emerged as a versatile
tool for creating new data instances that closely resemble a given training set~\cite{Goodfellow2014}. This innovative
paradigm involves a two-player adversarial setup where a generative ANN strives to produce data instances that are
indistinguishable from the training set, while a discriminative network attempts to distinguish between the instances
generated by the generative network and the real data~\cite{Salimans2016}. The competing nature of these two networks 
drives the generative network to generate increasingly realistic data, pushing the boundaries of what is achievable 
with generative models. This technology has been instrumental across a broad range of scientific disciplines,
including physics, chemistry, and biology. In physics, GANs have been used for simulating complex systems and predicting
outcomes of experiments, with examples in high energy physics~\cite{Paganini2018}, condensed matter physics~\cite{Carrasquilla2019,Koch2022,Carracedo2022}, nanophotonics~\cite{Rho2019,Christensen2020},
and cosmology~\cite{Rodriguez2018}. In the field of chemistry, GANs have been harnessed to generate 
novel chemical structures and predict their properties~\cite{Anstine2023}, thereby accelerating the process of drug discovery and materials
design~\cite{Sanchez-Lengeling2018}. In biology, GANs have been employed in a variety of tasks, including protein engineering~\cite{Repecka2021} and generate biological imaging data~\cite{Goldsborough2017}. This myriad of applications demonstrate the significant potential of GANs in transforming scientific
research by providing a powerful tool for hypothesis generation, experimental design, and data augmentation where 
empirical data is scarce or expensive to obtain. Despite their significant success in image generation, the use of GANs has been mostly limited to this area. It would be highly desirable to see their application more widely spread in the generation of scientific numerical data.
 
In this work, we introduce a novel application of GANs for synthetic spectral data generation. This offers a solution 
to the data scarcity found in scientific contexts where collecting a significant amount of spectral signals is critical 
for a subsequent application of data-driven approaches. Such a scenario is common across a wide range of fields, 
including physics, chemistry, astronomy, biology, medicine, and geology. Here, we particularly focus 
on an illustrative problem in the research area of near-field radiative heat transfer, involving a multilayered 
hyperbolic metamaterial. We explore the use of a Conditional Wasserstein GAN (CWGAN) for data augmentation and 
investigate its impact on the predictive capabilities of a feed-forward neural network (FFNN). We find that the successful production of spectral data requires two main changes to standard GANs. Firstly, the implementation of Wasserstein GANs (WGANs) is necessary to counteract mode collapse, and secondly, these WGANs need to be conditioned to yield accurate labels for the generated data. We demonstrate that a simple FFNN, when augmented with data produced by a CWGAN, notably improves its performance under conditions of data scarcity. This underscores the intrinsic value of CWGAN data augmentation, not just as a means to expand datasets. Furthermore, we illustrate that CWGANs have the ability to serve as efficient surrogate model in low-data regimes. Overall, this research work contributes to highlighting generative AI algorithms'
potential in applications extending beyond the conventional realm of image generation. We also anticipate that our
findings will contribute to advancing the understanding and application of generative AI algorithms in data-limited
scientific contexts.

This work is organized as follows.~In Section~\ref{sec-GANs}, we review the fundamentals of the primary generative adversarial frameworks underlying this work. Section~\ref{sec-system} discusses the basic principles of the particular physical problem we utilize to exemplify our approach. In Section~\ref{sec-results}, we present and discuss the results obtained from implementing the generative adversarial method detailed in Section 2 to the specific example problem outlined in Section 3. Finally, in Section~\ref{sec-conclusions} we sum up the conclusions of this work.

\section{Generative adversarial route to synthetic spectral data generation} \label{sec-GANs}

\begin{figure}
\includegraphics[width=\textwidth]{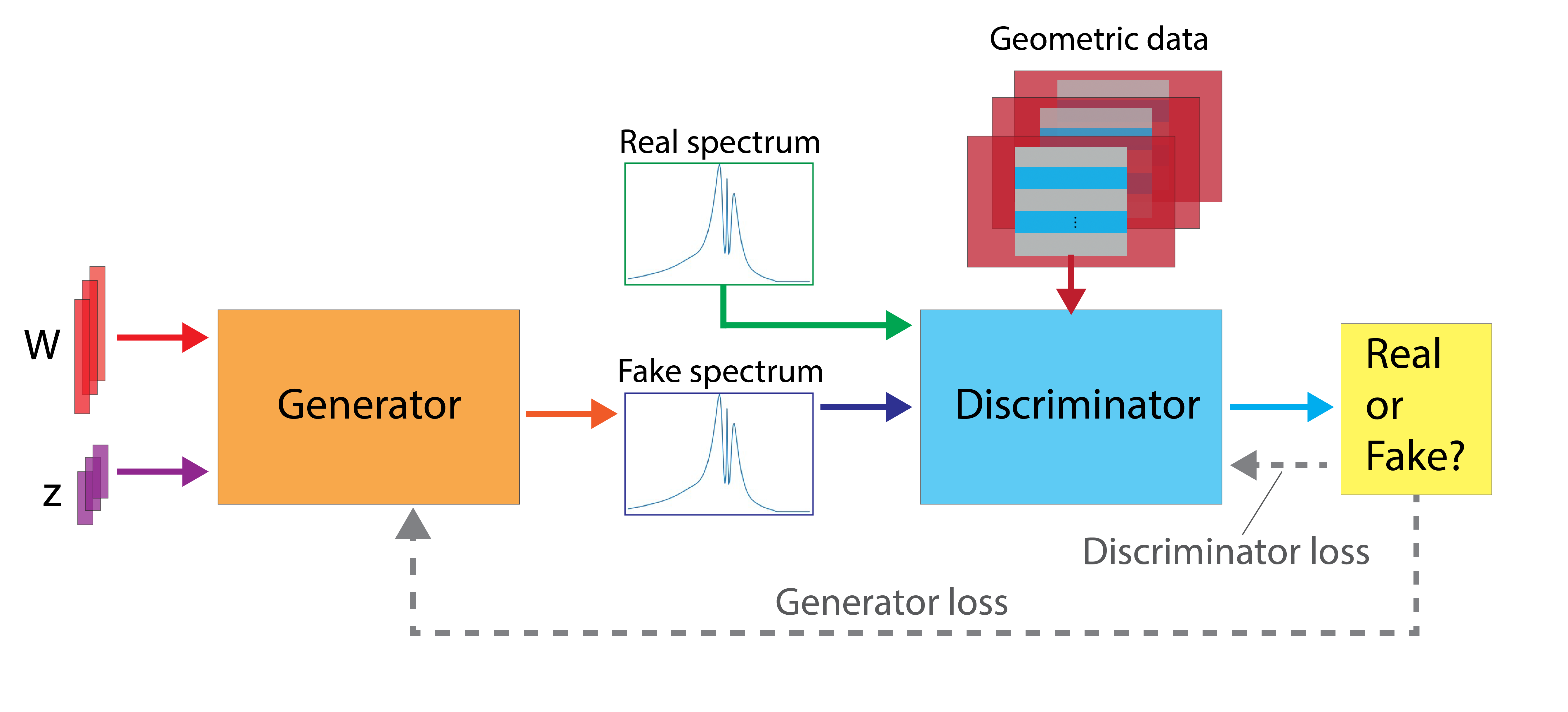}
\caption{Schematic representation of a Conditional Generative Adversarial Network (CGAN) employed to produce synthetic
spectral data. The process begins with the combination of geometrical data (\emph{\bf{W}}), and a random vector
(\emph{\bf{z}}), both of which are fed into the generator. The generator's role is to create a synthetic, or 'fake', 
spectrum from these inputs. The generated spectrum is then introduced to the discriminator, along with the real spectra 
and the geometrical data. The discriminator's task is to discern between the fake and real spectra. The conclusions 
drawn by the discriminator are subsequently used to guide the training of both the generator and discriminator 
networks, enhancing their ability to generate realistic spectra and to distinguish between real and synthetic data,
respectively.}
\label{fig-CGAN}
\end{figure}

In this Section, we review the basics of the three main generative adversarial frameworks that form the basis of this 
work, namely, GANs, Wasserstein GANs (WGANs) and Conditional WGANs (CWGANs). Here, we assume that the origin of the 
real spectra used in these approaches is completely general, coming from any from physical, chemical or biological 
process (in the following Sections we discuss the application to a particular problem within the realm of near-field
radiative heat transfer). Figure~\ref{fig-CGAN} shows a schematic representation of the underlying architecture of a 
general GAN algorithm. It comprises two main parts: a generator network and a discriminator network (represented as 
orange and blue rectangles, respectively, in Fig.~\ref{fig-CGAN}). The generator and the discriminator are two 
interconnected networks in the GAN system. The generator generally begins with random noise ($\bf{z}$ in Fig.~\ref{fig-CGAN}) and 
uses it to create new data samples (new spectra in our application). On the other hand, the discriminator takes these spectra
and calculates the likelihood that each one originates from the actual training data set. The two networks have competing
objectives. The generator's goal is to create spectra that perfectly mirrors the distribution of the training data. 
In doing so, it aims at generating spectral data so convincingly authentic that the discriminator cannot tell it apart 
from the real training data. Meanwhile, the discriminator aims at distinguishing between the actual training data and 
the data fabricated by the generator. Both networks are trained together in a competition until a Nash-type equilibrium 
is reached and the training process ends~\cite{Goodfellow2014}. At that stage, the generator is producing spectra that 
the discriminator can no longer reliably classify as `real' or `fake', signaling the end of the training process. 

Mathematically, the GAN configuration can be formulated as a minimization problem of the generator and discriminator 
loss functions, suitably written in terms of differentiable functions representing the discriminator and the generator
network models, $D({\bf x}; \bm{\theta}_d)$ and $G({\bf z}; \bm{\theta}_g)$, respectively ($\bm{\theta}_d$ and
$\bm{\theta}_g$ are the parameters of the corresponding ANNs ---in what follows, for the sake of clarity, we do not include 
this dependence in the equations). We start by focusing on the loss function of the discriminator, $L_D(G,D)$, which can 
be written as~\cite{Goodfellow2014}
\begin{equation} \label{eq-LD}
\centering
L_D(G,D) = - \mathbb{E}_{{\bf x}\sim p_{\mathrm{data}}({\bf x})}[\log(D({\bf x}))] - 
\nonumber \mathbb{E}_{{\bf z} \sim p_{z}({\bf z})}[\log(1-D(G(\bf{z})))] ,
\end{equation}
where $\mathbb{E}$ stands for expectation over either the training samples, $\bf{x}$, or the input noise variables, 
$\bf{z}$~(characterized respectively by a probability distribution $p_{\mathrm{data}}({\bf x})$ and a prior distribution 
$p_{z}({\bf z})$). From Eq.~(\ref{eq-LD}), and considering that $D$ outputs a single scalar representing a probability, 
we can see that the training of the discriminator is trying to minimize the likelihood of mistaking a real sample for a 
fake one or a fake sample for a real one (first and second terms, respectively, of the right-hand side of 
Eq.~(\ref{eq-LD})).

However, there are significant performance issues associated to this original choice of the loss function $L_D(G,D)$,
including the difficulty to reach the above-mentioned Nash equilibrium state (each network is updated independently, 
and given the competitive nature of the generator and the discriminator, in general, there is no clear point to stop 
the training) or the so-called \emph{mode collapse} (arising when a network fails to generalize accurately to all regions 
of the training data distribution). To address these issues, a different strategy to train GANs was introduced, the 
so-called \emph{Wasserstein Generative Adversarial Networks} (WGANs)~\cite{Arjovsky2017Jan}. One of the main 
modifications of WGANs with respect to the original GAN architecture is the presence of a critic model, $C({\bf x})$,
instead of the discriminator model. Importantly, $C({\bf x})$ outputs a score instead of a probability, which, in turn,
allows us to define a new loss function for the critic, $L_C(G,C)$, given by
\begin{equation} \label{eq-LC-v1}
\centering
     L_C(G,C) = -\mathbb{E}_{x\sim p_{\mathrm{data}}({\bf x})}[C({\bf x})] + 
     \mathbb{E}_{z\sim p_{z}({\bf z})}[C(G({\bf z}))] .
\end{equation}
This loss function summarizes well some of the main advantages of the WGANs. WGANs help address the training challenges 
of GANs by using the Wasserstein distance (or Earth Mover's distance) as the loss function instead of the Jensen-Shannon
divergence used in original GANs~\cite{Arjovsky2017Jan}. In addition, it provides a meaningful loss metric, i.e., the 
value of the critic in WGANs provides a meaningful measure of the distance between the real and generated data
distributions. This is unlike the original GANs where the discriminator's output does not correlate well with the quality 
of the generated samples.

A pivotal aspect of WGANs is the need for the Critic to operate within the set of the so-called \emph{1-Lipschitz} 
functions, a critical component of the model~\cite{Arjovsky2017Jan}. Lipschitz functions are mathematical functions
possessing a property where there exists a real-valued constant such that, for every pair of points, the absolute 
difference in function values can be bounded by this constant times the absolute difference of input values. When 
this constant is 1, the functions are known as 1-Lipschitz. The 1-Lipschitz constraint is crucial as it bounds how 
much a function's output can change with small variations in input, ensuring the function does not change too abruptly. 
In WGANs, this is vital to ensure that the Critic provides meaningful and stable gradients for the generator to learn from,
facilitating a more reliable learning process. Originally, weight clipping was proposed to enforce this 1-Lipschitz
condition. However, this sometimes resulted in convergence failure~\cite{Arjovsky2017Jan}. To counter these issues, 
a more robust technique known as the \emph{gradient penalty} was introduced. This method involves adding a loss term 
to maintain the L2 norm (a measure of the vector length of parameters or weights) of the Critic close to a value of
1~\cite{Gulrajani2017Mar}. This approach assists in keeping the Critic's function within the 1-Lipschitz constraint,
enhancing the stability and performance of WGANs. Incorporating these improvements, the full loss for the critic in a 
WGAN now reads
\begin{eqnarray} \label{eq-loss-discr}
L_C(G,C) & = & -\mathbb{E}_{{\bf x} \sim p_{\mathrm{data}}({\bf x})}[C({\bf x})] + 
\mathbb{E}_{{\bf z}\sim p_{z}({\bf z})}[C(G({\bf z}))] + \nonumber \\
 & & \lambda \: \mathbb{E}_{{\bf \hat{x}}\sim p_{\hat{x}}}
 \left[( \vert\vert\nabla_{\hat{x}}C({\bf \hat{x}})\vert\vert_2-1)^2 \right] ,
\end{eqnarray}
where $\vert\vert\cdot\vert\vert_2$ is the L2 norm and $\lambda$ is a weight parameter for the gradient penalty 
(throughout this work we assume $\lambda = 10$, as pointed out in Ref.~\cite{Gulrajani2017Mar}), and $\hat{x} = x + 
\alpha(G(z)-x)$ is an interpolated point between a real sample and generated sample on which to calculate the gradient, 
with $\alpha \in [0,1]$. As a final remark, proper training of the WGAN requires the critic to be trained ahead of the
generator, so that for each training step of the generator the critic is updated $n_{\rm train}$ times 
(following~\cite{Arjovsky2017Jan}, we chose $n_{\rm train} = 5$ in all our models).

As for the generator loss function, $L_G(G,C)$, an important aspect to realize is that for synthetic spectral data
generation our focus is on a regression problem. This implies that we need a greater control over the output than 
that obtained just by acquiring a random but realistic sample. It is therefore crucial to ensure that the generated 
data accurately corresponds to the correct system parameters that yield that response. To achieve this, we need to 
\emph{condition} the WGAN, leading to the creation of a Conditional WGAN (CWGAN)~\cite{Mirza2014Nov}. In this work, 
we will implement that by adding an extra loss term quantifying the Mean Absolute Error (MAE) between the conditioned
generated example and the training example corresponding to the \emph{ground truth} of the condition. Accounting for 
this conditioning, $L_G(G,C)$ can be expressed as
\begin{equation} \label{eq-LG}
L_G(G,C) = \mathbb{E}_{{\bf z} \sim p_z({\bf z}), \: {\bf x} \sim p_{data}({\bf x})} \left[ \left| {\bf x} - G({\bf z}\vert {\bf W})\right| \right] - 
\mathbb{E}_{{\bf z} \sim p_z({\bf z})}[C(G({\bf z}))] ,
\end{equation}
where ${\bf x}$ is the training example corresponding to the system parameters ${\bf W}$, and $G({\bf z}\vert {\bf W})$ 
is a generated example conditioned on the same parameters ${\bf W}$. The first term in the r.h.s. of Eq.~(\ref{eq-LG})
corresponds to the above discussed conditioning procedure, while the second term is associated to the coupling of the
generator and the critic.  

\section{Illustrative problem: Near-field radiative heat transfer spectra in multilayer hyperbolic metamaterials} 
\label{sec-system}

In this Section we provide an overview of the fundamentals of the specific problem we use to illustrate the proposed
approach. We have chosen a physical problem in the context of near-field radiative heat transfer involving multilayer 
hyperbolic metamaterials~\cite{Garcia-Esteban2021Dec}. Despite the specific character of this class of systems, the 
chosen problem can be considered both representative of the types of problems that our approach can address effectively 
and complex enough to showcase the versatility of our method.  

One of the major advances in recent years in the field of thermal radiation has been the experimental confirmation 
that the limit set by Stefan-Boltzmann's law for the radiative heat transfer between two bodies can be largely overcome 
by bringing them sufficiently close \cite{Polder1971Nov}. This phenomenon is possible because in the near-field regime,
i.e., when the separation between two bodies is smaller than the thermal wavelength $\lambda_{\rm Th}$ ($\sim$10 $\mu$m 
at room temperature), radiative heat can also be transferred via evanescent waves (or photon tunneling). This new 
contribution is not taken into account in Stefan-Boltzmann's law and dominates the near-field radiative heat transfer
(NFRHT) for sufficiently small gaps or separations~\cite{Song2015May,Cuevas2018Oct,Biehs2021Jun}. Among the different
strategies that have been recently proposed to further enhance NFRHT, one of the most popular ideas is based on the use 
of multiple surface modes that can naturally appear in multilayer structures. In this regard, a lot of attention has 
been devoted to multilayer systems where dielectric and metallic layers are alternated to give rise to the so-called
\emph{hyperbolic metamaterials}~\cite{Guo2012Sep,Biehs2012Sep,Guo2013Jun,Biehs2013Apr,Bright2014Jun,Miller2014Apr,
Biehs2017Feb,Iizuka2018Feb,Song2020Feb,Moncada-Villa2021Feb}. The hybridization of surface modes appearing in different
metal-dielectric interfaces have indeed been shown to lead to a great enhancement of the NFRHT, as compared to the case 
of two infinite parallel plates~\cite{Iizuka2018Feb}. 

\begin{figure}[t]
\includegraphics[width=\textwidth]{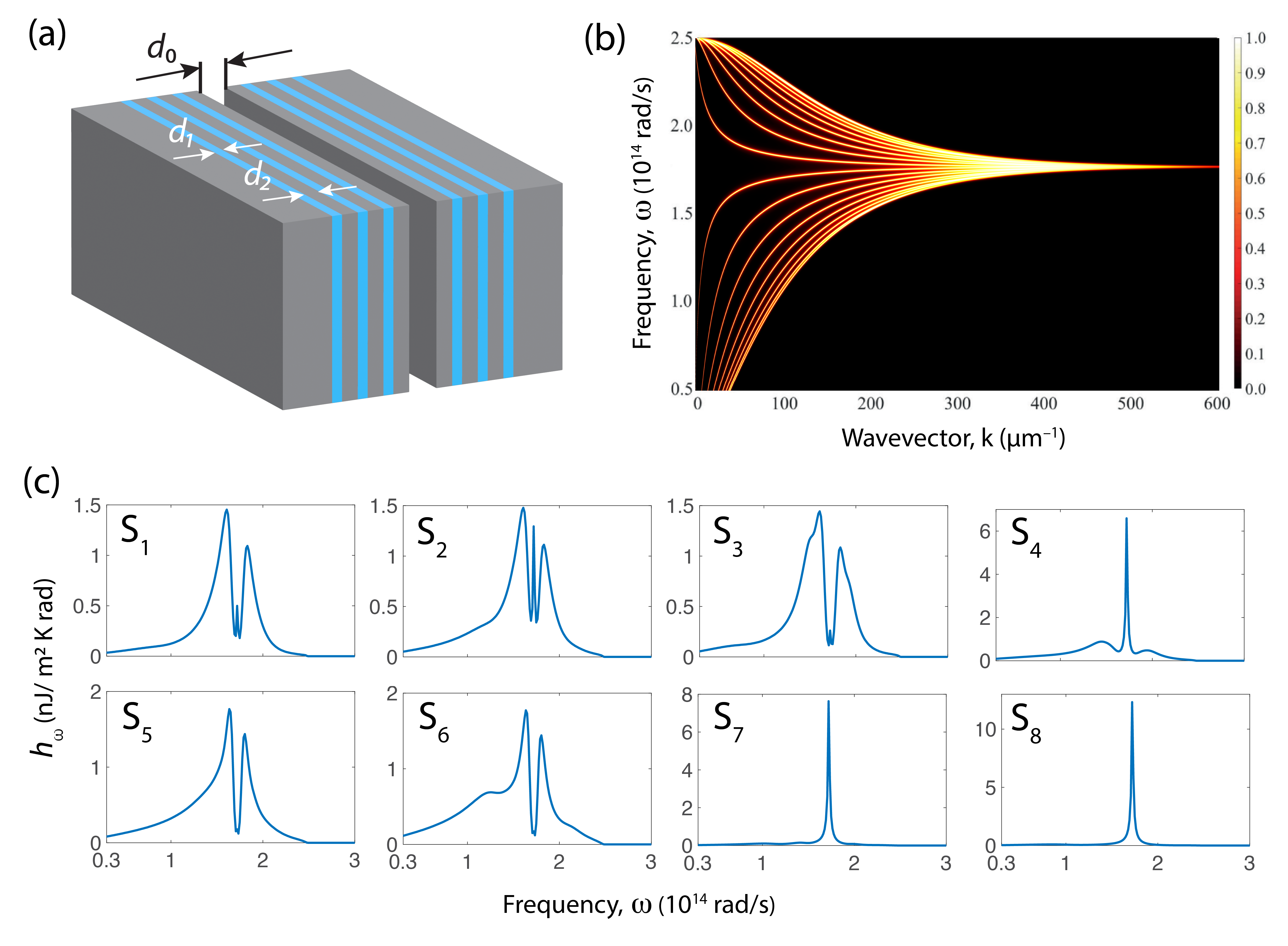}
\caption{(a) Schematic representation of the physical system under study. It features a multilayered hyperbolic 
metamaterial comprised of two identical structures, each alternating between metal (depicted in grey) and dielectric 
(blue) layers that extend infinitely. These two systems are separated by a distance $d_0 = 10$~nm. Every layer has a 
thickness $d_i$, and both systems are backed by a metallic substrate. (b) Transmission of evanescent waves as a function 
of the frequency ($\omega$) and the parallel wavevector ($k$), considering the case where there are 8 layers per system 
and both $d_i=d_0=10$~nm. The transmission pattern exhibits a series of lines nearing unity, resulting from the
hybridization of surface plasmon polaritons (SPPs) within the metallic layers. (c) Sample spectra of heat transfer
coefficients, $h_\omega$, for several representative combinations of layer thicknesses, which are specified in 
Table~1.}
\label{fig-draw}
\end{figure}

Following Ref.~\cite{Iizuka2018Feb}, we consider here the radiative heat transfer between two identical multilayer
structures separated by a gap $d_0$, as shown in Fig.~\ref{fig-draw}(a). Each body contains $N$ total layers alternating
between a metallic layer with a permittivity $\epsilon_\mathrm{m}$ and a lossless dielectric layer of permittivity
$\epsilon_\mathrm{d}$. The thickness of the layer $i$ is denoted by $d_i$ and it can take any value within a given 
range (to be specified below). While the dielectric layers will be set to vacuum ($\epsilon_\mathrm{d} =1$), the 
metallic layers will be described by a permittivity given by a Drude model: $\epsilon_\mathrm{m}(\omega) = 
\epsilon_{\infty} - \omega^2_p/[\omega (\omega + i \gamma)]$, where $\epsilon_{\infty}$ is the permittivity at 
infinite frequency, $\omega_p$ is the plasma frequency, and $\gamma$ de damping rate. From now on, we set 
$\epsilon_{\infty} = 1$, $\omega_p = 2.5 \times 10^{14}$ rad/s, and $\gamma = 1 \times 10^{12}$ rad/s (these 
parameters provide a surface plasmon frequency similar to the surface phonon-polariton frequency of the interface 
between SiC and vacuum).

We describe the radiative heat transfer within the framework of fluctuational electrodynamics 
\cite{Rytov1953,Rytov1989}, particularly focusing on the near-field regime. In this regime, the radiative heat transfer 
is dominated by TM- or $p$-polarized evanescent waves and the heat transfer coefficient (HTC) between the two bodies, 
i.e., the linear radiative thermal conductance per unit of area, is given by \cite{Basu2009Oct}
\begin{equation}
h = \frac{\partial}{\partial T} \int^{\infty}_0 \frac{d\omega}{2\pi} \: \Theta(\omega, T) 
\int^{\infty}_{\omega/c} \frac{dk}{2\pi} \, k \: \tau_p(\omega, k) ,
\end{equation}
where $T$ is temperature, $\Theta(\omega, T)= \hbar \omega/ (e^{\hbar \omega/ k_{\rm B} T} -1)$ is the mean thermal 
energy of a mode of frequency $\omega$, $k$ is the magnitude of the wave vector parallel to the surface planes, and
$\tau_p(\omega, k)$ is the transmission (between 0 and 1) of the $p$-polarized evanescent modes given by
\begin{equation}
 \tau_p(\omega, k) = \frac{4 \left[ \mbox{Im} \left\{ r_p(\omega,k) \right\} \right]^2 e^{-2q_0 d_0}}
 {| 1 - r_p(\omega,k)^2 e^{-2 q_0 d_0} |^2} .
\end{equation}
Here, $r_p(\omega,k)$ is the Fresnel reflection coefficient of the $p$-polarized evanescent waves from the vacuum 
to one of the bodies and $q_0 = \sqrt{k^2 - \omega^2/c^2}$ ($\omega/c < k$) is the wave vector component normal to 
the layers in vacuum. The Fresnel coefficient needs to be computed numerically and we have done it by using the 
scattering matrix method described in Ref.~\cite{Caballero2012Jun}. In our numerical calculations of the HTC we also 
took into account the contribution of $s$-polarized modes, but it turns out to be negligible for the gap sizes explored 
in this work.

Let us briefly recall that, as explained in Ref.~\cite{Iizuka2018Feb}, the interest in the NFRHT in these multilayer
structures resides in the fact that the heat exchange in this regime is dominated by surfaces modes that can be shaped 
by playing with the layer thicknesses. In the case of two parallel plates made of a Drude metal, the NFRHT is dominated 
by the two cavity surface modes resulting from the hybridization of the surface plasmon polaritons (SPPs) of the two 
metal-vacuum interfaces \cite{Iizuka2018Feb}. These two cavity modes give rise to two near-unity lines in the 
transmission function $\tau_p(\omega, k)$. Upon introducing more internal layers with appropriate thicknesses, one can 
have NFRHT contributions from surface states at multiple surfaces, as we illustrate in Fig.~\ref{fig-draw}(b) for the case of $N=8$
layers (4 metallic and 4 dielectric layers), separated by a vacuum gap of $d_0 = 10$ nm at $T=300$ K. As shown in 
Ref.~\cite{Iizuka2018Feb}, the contribution of these additional surface states originating from internal layers can 
lead to a great enhancement of the NFRHT as compared to the bulk system (two parallel plates) in a wide range of gap 
values.

\begin{table}
\caption{Thicknesses combinations of the representative samples of the spectral heat transfer coefficient, 
$h_\omega$, shown in Fig.~\ref{fig-draw}(c).}
\begin{indented}
\lineup
\item[]\begin{tabular}{@{}c*{8}{r}}
\br   
&\centre{8}{Layer thicknesses (nm)} \\
\ns
&\crule{8}  \\
Label&$d_1$&$d_2$&$d_3$&$d_4$&$d_5$&$d_6$&$d_7$&$d_8$ \\ 
\mr
$S_1$& 5.0 & 20.0 & 20.0 & 5.0 & 20.0 & 5.0 & 5.0 & 20.0 \\ 
$S_2$& 5.0 & 12.5 & 12.5 & 5.0 & 12.5 & 5.0 & 5.0 & 12.5 \\ 
$S_3$& 5.0 & 12.5 & 5.0 & 5.0 & 5.0 & 20.0 & 12.5 & 5.0 \\  
$S_4$& 5.0 & 5.0 & 12.5 & 5.0 & 5.0 & 5.0 & 20.0 & 12.5 \\ 
$S_5$& 5.0 & 5.0 & 5.0 & 20.0 & 12.5 & 12.5 & 5.0 & 20.0 \\ 
$S_6$& 5.0 & 5.0 & 5.0 & 12.5 & 5.0 & 20.0 & 5.0 & 12.5 \\ 
$S_7$& 20.0 & 5.0 & 20.0 & 12.5 & 20.0 & 5.0 & 12.5 & 20.0 \\  
$S_8$& 12.5 & 12.5 & 5.0 & 12.5 & 5.0 & 5.0 & 12.5 & 5.0 \\ 
\br
\end{tabular}
\end{indented}
\end{table}

Of special interest for this work is the spectral HTC, $h_\omega$, defined as the HTC per unit of frequency: $h = 
\int^{\infty}_0 h_\omega \: d\omega$. To create the initial dataset of real spectra (which later on will be augmented
by our generative adversarial approach), we apply the above-described theoretical framework to compute a total of 6,561
$h_\omega$ spectra. The thicknesses $d_i$ of each layer were varied between 5 and 20 nm, and every spectrum contains 200
frequency points in the range $\omega \in [0.3, 3] \times 10^{14}$ rad/s. As discussed in the next Section, that dataset
of spectra will be split in different proportions to become training and test sets. Figure~\ref{fig-draw}(c) shows 
several representative samples of $h_\omega$ spectra,  corresponding to the following thicknesses combinations listed 
in Table~1.

The spectra displayed in Fig.~\ref{fig-draw}(c) show the broad variety of spectral features that can be obtained 
from the studied system (from double broad peaks with very narrow resonances in between, to single narrow peaks, or 
to two resonant peaks separated by a gap). This set of spectra essentially serves as a comprehensive blueprint for 
the whole adversarial approach, guiding it on the characteristics and features that should be exhibited in the 
synthetic data. Hence, this allows the proposed approach to capture a wider range of underlying patterns and 
relationships, which in turn allows it to generate a more realistic and diverse array of synthetic spectral data.

\section{Results and discussion} \label{sec-results}

\begin{figure}[t]
\includegraphics[width=\textwidth]{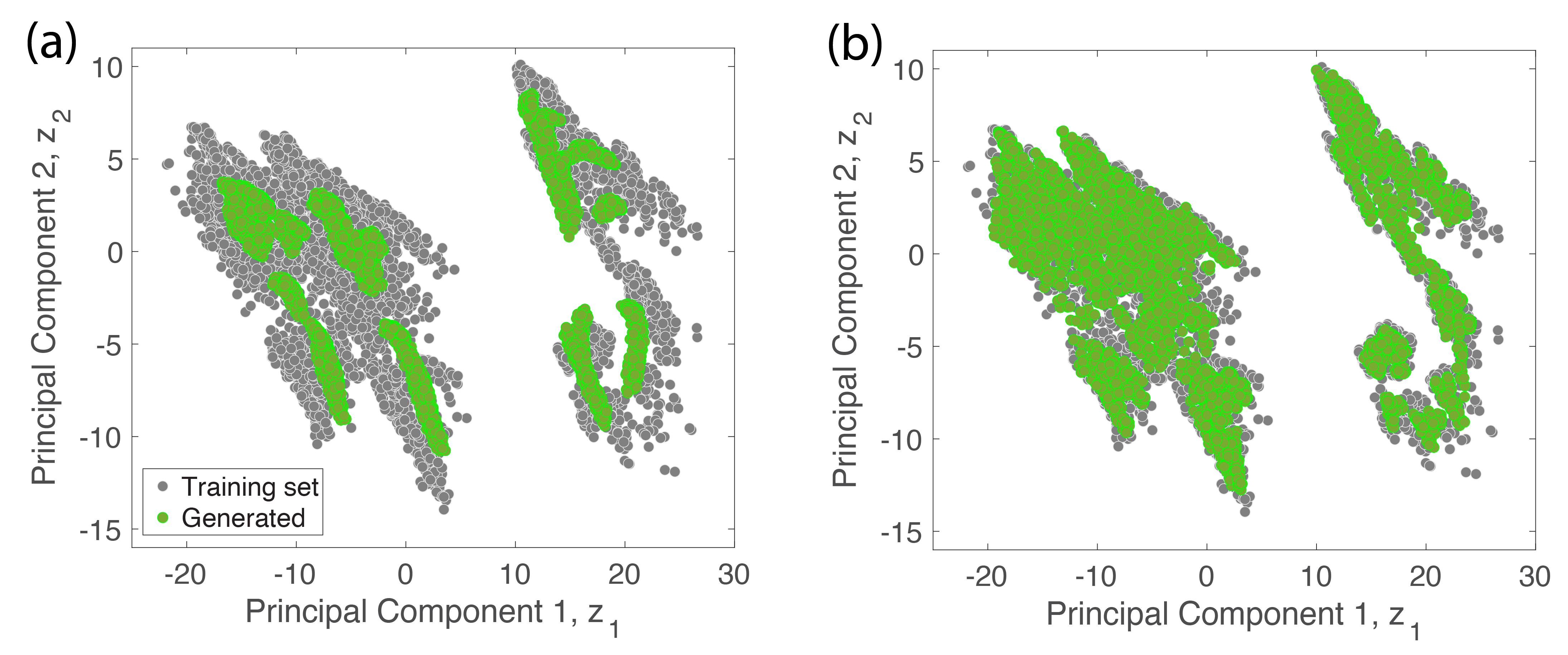}
\caption{Reproduction of the training set using (a) a CGAN, showing in grey the training set and in green the CGAN
reproduction of the set; and (b) the CWGAN, using the same color scheme. The results  were obtained by applying PCA to the
corresponding spectra, and plotting the first two components, $z_1$ and $z_2$ (with a $>90\%$ reproduction rate).}
\label{fig-PCA}
\end{figure}

We proceed in this Section to report on the results obtained when applying the generative adversarial approach 
summarized in Section~\ref{sec-GANs} to the specific illustrative problem described in Section~\ref{sec-system}. 
We begin by providing a strong quantitative justification of the necessity of using a trained CWGAN for this problem,
instead of a plain CGAN (Conditional GAN ---the details of the specific architecture used in each case are provided below). 
Figures~\ref{fig-PCA}(a) and ~\ref{fig-PCA}(b) show, respectively, the ability to reproduce the training set of a 
trained CGAN and a trained CWGAN projecting the data on two dimensions via a Principal Component 
Analysis (PCA)~\cite{Jolliffebook}, which retains most of the training data structure due to its $>90\%$ reproduction 
rate ($RR$).

The PCA calculations shown in Fig.~\ref{fig-PCA} were done as follows. First, we performed a singular value 
decomposition (SVD) of the covariance matrix $\Sigma$: $\left[{\bf U},{\bf S},
{\bf V} \right] = S\,V\,D(\Sigma)$ with $\Sigma = \frac{1}{m}X^TX$~\cite{JamesWittemHastieTibshiranibook}. Here, \textit{U} and \textit{V} are unitary matrices,
\textit{S} is the singular value matrix, \textit{m} is the total number of data examples and \textit{X} is the data 
matrix, containing in each column one data example \textbf{x}. To obtain the reduced 2-dimensional (2D) representation of 
the data, we calculated a reduced matrix $\mathrm{U}_{\mathrm{reduced}}$ retaining the first 2 columns of the U matrix
obtained from the SVD, and use it as the projection matrix, $\bf\hat{x} = \mathrm{U}^T_{\mathrm{reduced}}X$, where
$\bf\hat{x}$ is the 2-dimensional representation of the data. The reproduction rate ($RR$) of the whole 2D-PCA analysis 
is then defined as the ratio of the first 2 singular values over all the $N$ singular values obtained:
\begin{equation}
\mathrm{RR} = \frac{\sum^2_{i=1}S_{ii}}{\sum^N_{i=1}S_{ii}} 
\end{equation}

The Conditional GAN and the Conditional WGAN share the same architectural design, the specifics of which are detailed
further on. Both networks underwent identical training conditions, with the complete spectral dataset partitioned into 
80\% for training and 20\% for validation. The key distinguishing factor lies in their respective loss functions 
($L_D(G,D)$ for CGAN and $L_C(G,C)$ for CWGAN). The CGAN employs Eq.~(\ref{eq-LD}) along with a generator loss which is the sum of the first term in the r.h.s. of Eq.~(\ref{eq-LG}) (the conditional term) and the second term in the r.h.s.~of Eq.~(\ref{eq-LD}) with opposite sign. Meanwhile, the CWGAN utilizes the loss defined in Eq.~(\ref{eq-LG}) for the generator and Eq.~(\ref{eq-loss-discr}) for the critic. Notably, as depicted in Fig.~\ref{fig-PCA}(a), 
the CGAN manifests evident signs of mode collapse, rendering it unable to reproduce most examples beyond the principal
cluster structures in the PCA. In contrast, the CWGAN shows the capability of replicating
most of the complexities within the training set. These results can be considered as a novel illustration, in the context 
of synthetic generation of spectral data, of the key role played by Wasserstein's loss function to create 
more robust generative adversarial approaches.

\begin{figure*}[t]
\centering
\includegraphics[width=14cm]{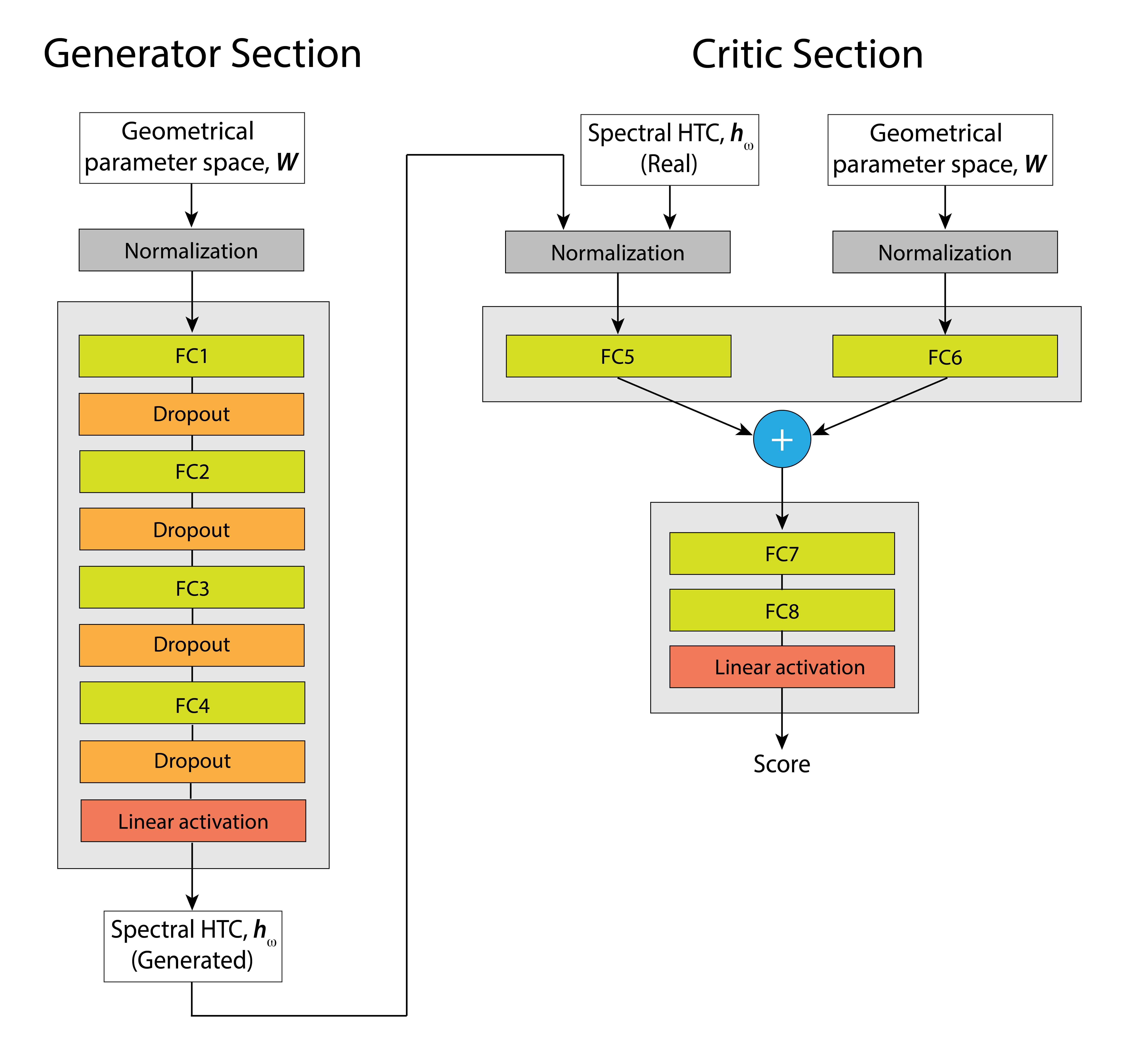}
\caption{Schematics of the architecture used for the Conditional Wasserstein Generative Adversarial Network (CWGAN) 
employed in this work. Both the generator and critic sections are shown in the left and right parts, respectively, of the
figure. As shown, the generator accepts geometrical parameters as input and produces a heat transfer coefficient 
(HTC) spectrum as output. It is constructed of four fully connected layers (FCs) (green blocks), dropout layers 
placed after each FC layer (orange blocks) and a final linear activation layer (red block). The critic section simultaneously
 accepts  HTC spectra produced by the generator, as well as \emph{true} spectra, and the geometrical parameters as 
inputs, providing a scalar score as output. The critic model consists of a unique FC layer for each parallel input branch followed 
by a concatenation layer that combines information from both input branches into a single vector. Afterwards, two more 
FC layers and a final linear activation layer produce the score. In both the generator and critic sections, normalization of the input 
data is applied as a preprocessing step (grey blocks).}
\label{fig-nets}
\end{figure*}

Figure~\ref{fig-nets} summarizes the specific CWGAN architecture we have found as most efficient for the studied problem.
The generator (left side of Fig.~\ref{fig-nets}) is composed of 4 hidden fully connected layers of an increasing number of neurons 
in each layer (we consider 50, 100, 150, and 200 neurons for the four layers ---represented as green blocks, labeled as FC1--FC4, in 
Fig.~\ref{fig-nets}). The generator model takes the condition as input (the 8 values for the thicknesses of the 
layers) and returns a generated $h_\omega$ spectrum (sampled by 200 frequencies). Consistently with the results in 
Refs.~\cite{Isola2016Nov,Ahmed2021Sep}, we found that we did not need to feed a random data distribution \textbf{z} into 
the generator for operation: 20$\%$ dropout layers between all layers of the generator (i.e., connections between two
consecutive neurons are dropped with a 20\% chance) provide enough variability for this task (the dropout operations 
are represented as orange blocks in Fig.~\ref{fig-nets}). On the other hand, the critic (right side of Fig.~\ref{fig-nets}) 
takes both a sample spectrum and the condition (either generated or from the training distribution) in parallel 
processing lines of a single hidden layer (with 150 and 50 neurons, respectively ---FC5 and FC6 in Fig.~\ref{fig-nets}). Then, it concatenates and processes the information through two additional 
hidden layers (FC7 and FC8 in Fig.~\ref{fig-nets}, with 100 and 50 neurons, respectively) to output a single number, the score. All hidden layers feature a scaled exponential linear unit (selu)
activation function. In all models discussed in this work, as a pre-processing step, we will also calculate the 
logarithm of the spectra, subtract the mean of both the input parameters and the spectra, and divide both the system
parameters and the spectra by the standard deviation (this normalization operation is represented by grey blocks 
in Fig.~\ref{fig-nets}). Finally in both the generator and the critic a final linear activation layer is added to ensure the output 
has the correct size (red blocks in Fig.~\ref{fig-nets}).

\begin{figure*}[t]
\centering
\includegraphics[width=10cm,clip]{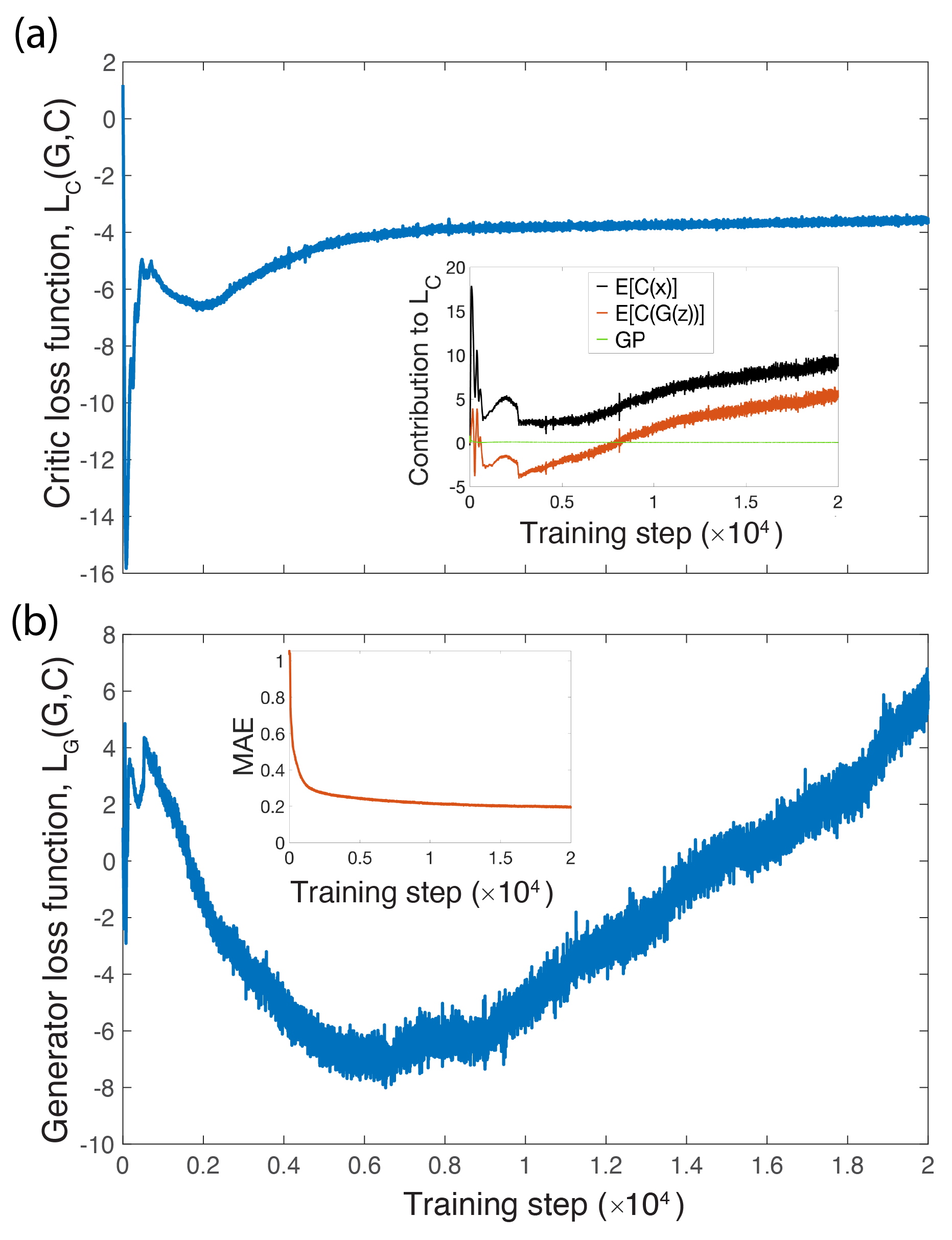}
\caption{Evolution of the critic and generator loss functions during the training process ($L_C(G,C)$ and 
$L_G(G,C)$, panels (a) and (b), respectively). Inset of panel (a) shows dependence on the training steps of three
contributions of $L_C(G,C)$ (see Eq.~\ref{eq-loss-discr} in the main text). Inset of panel (b) corresponds to 
evolution with the training step of the mean absolute error (MAE) associated to the synthetic spectra produced 
by the generator. Notice that in our framework the critic evaluates the loss 5 times per generator loss evaluation.}
\label{fig-loss}
\end{figure*}

We focus now on analyzing the evolution with the training steps of the loss functions of the discriminator and the 
generator ($L_C(G,C)$ and $L_G(G,C)$, respectively), as obtained by training the architecture displayed in 
Fig.~\ref{fig-nets} with the 6,561 $h_\omega$ spectra described in Section~\ref{sec-system}. Monitoring the loss 
functions underlying our model during the training process is critical to understanding, diagnosing, and improving 
the whole generative adversarial framework studied in this work. The obtained numerical results are summarized in
Fig.~\ref{fig-loss} (panel (a) corresponds to $L_C(G,C)$, while panel (b) displays the results for $L_G(G,C)$). 
As seen in Fig.~\ref{fig-loss}(a), $L_C(G,C)$ initially displays a transient behavior based on a fast drop in 
value followed by a sudden increase and an oscillation (around $10^3$ training steps), until it reaches at 
stationary value at approximately $10^4$ training steps. To get additional insight into the numerical origin of this
evolution, the inset of Fig.~\ref{fig-loss}(a) shows the dependence with the training steps of the three different 
terms forming $L_C(G,C)$, namely, $\mathbb{E}_{{\bf x} \sim p_{\mathrm{data}}({\bf x})}[C({\bf x})]$, 
$\mathbb{E}_{{\bf z}\sim p_{z}({\bf z})}[C(G({\bf z}))]$, and the gradient penalty, $\lambda \: 
\mathbb{E}_{{\bf \hat{x}}\sim p_{\hat{x}}} \left[( \vert\vert\nabla_{\hat{x}}C({\bf \hat{x}})\vert\vert_2-1)^2 
\right]$ (see Eq.~\ref{eq-loss-discr}). As observed, $L_C(G,C)$ is dominated by the expected values of the score 
value produced by the critic model for both the true training samples, (${\bf x}$), and the samples fabricated 
by the generator, $G({\bf z})$ (black and red lines, respectively, in inset of of Fig.~\ref{fig-loss}(a)  ---the green 
line corresponds to the gradient penalty term). A remarkable aspect of the overall evolution of both expected value
contributions to $L_C(G,C)$ is the fact that, despite their complicated transient behavior, they lock their difference in values after around $10^4$ training steps. This in turn allows $L_C(G,C)$ to reach a stationary value 
(notice the difference in sign between both terms in Eq.~\ref{eq-loss-discr}), which marks the completion of the 
learning process by the critic.

Figure~\ref{fig-loss}(b) shows the results for the evolution of the generator loss function, $L_G(G,C)$, during 
the training process. The inset of Fig.~\ref{fig-loss}(b) displays the corresponding evolution of the MAE 
contribution to $L_G(G,C)$ (i.e., the evolution of the first term in the r.h.s.\ of Eq.~\ref{eq-LG}). As deduced 
by comparing the values reached by the MAE with those of $L_G(G,C)$, the generator loss function is dominated 
by the expected value term, $\mathbb{E}_{{\bf z}\sim p_{z}({\bf z})}[C(G({\bf z}))]$ (second term in the r.h.s.\ 
of Eq.~\ref{eq-LG}). As described above, for large values of the training step ($\gtrsim 10^4$), this latter 
term necessarily follows its counterpart term in the critic loss function (so, as also pointed out above, 
their difference is maintained and a stationary value of $L_C(G,C)$ is reached). Therefore, since we do not expect
$L_G(G,C)$ to reach a stationary value, MAE becomes the key magnitude to monitor the quality of the learning 
process of the generator (note that that this is also in accordance with the overall goal of the proposed 
application: creating artificial spectral samples indistinguishable from the real ones). Indeed, as shown in 
inset of Fig.~\ref{fig-loss}(b), the MAE computed for the studied problem converges to values $<0.2$ for the 
larger training step values considered in our calculations.

Next, we proceed to quantify our model's proficiency in reproducing the spectral heat transfer coefficient (HTC) 
of the studied physical system, based on the model architecture illustrated in Fig.~\ref{fig-nets}. For this analysis, 
our focus is on two distinctive metrics, concentrating on two different aspects of the spectra. The first, 
the per-point relative mean error ($L_{\mathrm{point}}$), gauges the model's capability to accurately represent 
each point in the spectrum through a relative error analysis. The second, the integral relative mean error
($L_{\mathrm{integ}}$), is predominantly influenced by the model's accuracy in reproducing the primary spectral
characteristics, notably the resonances. Their respective definitions are as follows:
\begin{eqnarray}
L_{\mathrm{point}} = \frac{1}{N\: m}\sum_{i=1}^N \sum_{j=1}^m  \frac{y^i_j - \hat{y}_j^i}{y^i_j} , \\
L_{\mathrm{integ}} = \frac{1}{N}\sum_{i=1}^N \frac{ \int (y^i - \hat{y}^i)\: d\omega}{\int y^i \: d\omega } ,
\end{eqnarray}
where \textit{N} is the number of spectra, \textit{m} the number of points in each spectrum, $y^i$ is a data 
example and $\hat{y}^i$ is the corresponding generated example by the model. Note that we undo all pre-processing
operations to perform these calculations, and that all integrals are performed over frequency points.

We start our analysis by establishing a baseline outcome for comparison using a simple FFNN, comprised of five hidden layers, each hosting 200 neurons,
characterized by a selu activation function, and a final linear activation layer. An analogous network has been previously shown 
to effectively model this identical physical system given ample data~\cite{Garcia-Esteban2021Dec} ---consequently, we
anticipate that this particular network will serve as a suitable benchmark. Another important aspect is the apparent similarities between the generator and the FFNN. We aim at demonstrating that this specific simple design, 
given a sufficiently large dataset, can accurately replicate the spectral characteristics inherent to the type 
of systems under study, without resorting to any data augmentation technique. To accomplish this, we split the 
original dataset, which consists of 6,561 $h_\omega$ spectra, allocating 80\% for training set and the remaining 
20\% for the validation set. Our findings reveal that after 50,000 training iterations (epochs), and using the 
Adam optimization algorithm with a steady learning rate of $3\times10^{-4}$, this simple neural network is 
proficient in replicating the system's spectra, reaching values of $L_{\mathrm{point}}$ and  $L_{\mathrm{integ}}$ 
for the validation set of $3.61\%$ and $1.45\%$, respectively (note that these results are comparable to those 
found in Ref.~\cite{Garcia-Esteban2021Dec}).

Then we proceed to assess the CWGAN capability as numerical engine for spectral data augmentation. To do that, 
we follow a two-step approach. First, once the CWGAN has been trained, we use it to generate a number of additional 
spectra to include on the training set. Second, we retrain the above-described FFNN with this new data set to create 
an \emph{augmented} FFNN. In this work, we generate a total of 10,000 new spectra for this data augmentation process, which are added to the training set. 
This particular amount of additional spectra was chosen after observing that our results are converged when adding 
to the original dataset a number of extra spectra in the range 5,000 -- 10,000. Note that, once the generator has 
been trained, this augmentation of the original dataset is computationally inexpensive. 

Following this approach, first, we found that when using the training/validation split considered until now 
(80\%-20\%), the FFNN and the augmented FFNN have similar performance, both in terms of per-point relative mean 
accuracy and integral relative mean error (respective error values of approximately 
 $3.6\%$ and $1.5\%$ are 
obtained for both models). This suggests that, as anticipated, using a sufficiently large dataset minimizes the 
impact of augmenting the original dataset, making any data augmentation barely noticeable. However, that conclusion 
changes when a different data scenario is considered. To modify the amount of data available to the models, we 
increasingly reduced the size of the training set by transferring part of it to the validation set, and retrained
both the simple and augmented FFNNs from scratch (different values of the split training/validation were 
considered in the range 80\%-20\% -- 1\%-99\% ---note that the data augmentation is done separately for each split ratio).

\begin{figure*}[t]
\includegraphics[width=\textwidth]{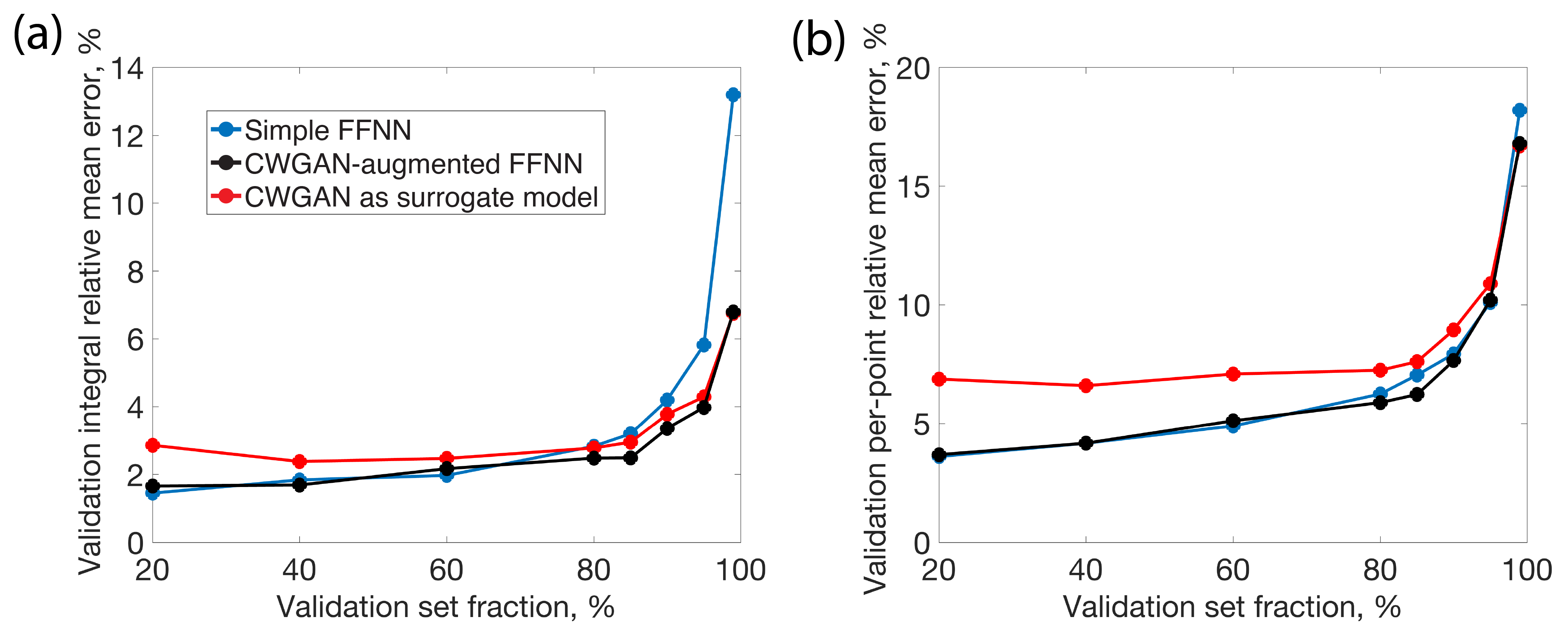}
\caption{(a) Evolution of the integral relative mean error for the validation set as a function of the validation set fraction of the total data. (b) Same as in (a), but now using the per-point relative mean error metric. In both (a) and (b), blue and black dots (lines) show the results for a simple FFNN and a CWGAN-augmented FFNN, respectively, whereas red dots (lines) correspond to the results obtained by using the CWGAN as surrogate model.}
\label{fig-lowdata}
\end{figure*}

Figure~\ref{fig-lowdata} summarizes the main results of this analysis using the two error metrics considered in 
this work (panels (a) and (b) correspond to $L_{\mathrm{integ}}$ and $L_{\mathrm{point}}$, respectively ---blue dots (lines) display the results for the simple FFNN, whereas black dots (lines) correspond to the CWGAN-augmented FFNN). As seen, 
both in terms of $L_{\mathrm{integ}}$ and $L_{\mathrm{point}}$, the simple FFNN and the augmented FFNN have similar
performance when the validation set fraction is smaller than approximately 70\% (we can therefore ascribe that 
interval to a sufficiently large training set scenario). However, in the case of $L_{\mathrm{integ}}$(Fig. ~\ref{fig-lowdata}(a)), further increasing of the validation set fraction (i.e., further reduction of the training set size) leads to the augmented FFNN to perform increasingly better with training data reduction than the simple FFNN . As observed, values of $L_{\mathrm{integ}}=13.2\%$ are achieved using a simple FFNN in 
the limit case of a validation set fraction of 99\%, whereas an augmented FFNN reduces that error close to a than 
a half ($L_{\mathrm{integ}}=6.8\%$). The performance improvement is not so dramatic in the case of the per-point
relative mean error (Fig.~\ref{fig-lowdata}(b)), but we still observe slight improvements in the low-data scenario 
(for the extreme case of a validation set fraction of 99\%, $L_{\mathrm{point}}=18.2\%$ is obtained for the simple 
FFNN, while the augmented FFNN leads to $L_{\mathrm{point}}=16.8\%$). 

Overall, the numerical results discussed above suggest that CWGAN data augmentation has intrinsic value beyond simply 
providing larger datasets. In particular, the reported results show that CWGAN data augmentation holds
value in creating synthetic data more adaptable and resilient to scenarios with limited data. To the best of our 
knowledge, this finding is reported here for the first time. We believe it could contribute to the development 
of more efficient models that require smaller datasets, through the synthetic generation capabilities offered 
by generative adversarial frameworks.

Finally, to conclude the present analysis, we compare the performance as surrogate model to generate $h_\omega$ 
spectra of the trained CWGAN with that of the simple and augmented FFNNs. This application arises from the conditioning
modification we introduced in this work to conventional GANs (see Eq.~\ref{eq-LG} and the corresponding discussion 
in Section~\ref{sec-GANs}). This conditioning allows us to accurately keep track of the geometrical parameters 
associated to each spectra created by the generator of our CWGAN model, so once the generator is trained, it can 
be decoupled from the whole architecture and used as a standalone surrogate model. The obtained numerical results are also displayed in Fig.~\ref{fig-lowdata} (red points and red lines corresponds to the results of a CWGAN used 
as surrogate model). As observed, for the integral relative mean error (Fig.~\ref{fig-lowdata}(a)) the CWGAN 
displays worse performance than both the simple and the augmented FFNNs when the validation set fraction is below 
70\% approximately (i.e., the data regime that we previously labeled as of sufficiently large training set scenario 
for both the simple and the augmented FFNNs). However, in the low-data regime (values of the validation set 
fraction larger than $70\%$) we observe that the CWGAN surrogate gradually improves the results of a simple FFNN, 
until fully reproducing the improvement reached by an augmented FFNN in the extreme case of 99\% validation set 
fraction. Regarding the per-point relative mean error (Fig.~\ref{fig-lowdata}(b)), the CWGAN surrogate displays a significantly worse performance in comparison to both the simple and the augmented FFNNs for the validation set fractions below 70\%, but from 
that point on, it starts converging to the results of an augmented FFNN in the low-data regime. Ultimately, this 
comparison between the surrogate role of all considered models reinforces our previous conclusion that the CWGAN shows higher 
resilience to the reduction of training data than the simple FFNN, becoming also in this context an efficient architecture in low-data 
scenarios.

\section{Conclusions} \label{sec-conclusions}

In this work we have studied the application of Generative Adversarial Networks (GANs) for synthetic spectral data 
generation, providing a solution to the data scarcity challenge pervasive in scientific domains where acquiring 
substantial spectral signals is of paramount importance. Our main focus has been an illustrative problem in the domain of near-field
radiative heat transfer involving a multilayered hyperbolic metamaterial. We have analyzed the use of a Conditional
Wasserstein GAN (CWGAN) for data augmentation and studied its influence on the predictive capacities of a feed-forward
neural network (FFNN). Our results reveal that generating spectral data effectively entails two main modifications to traditional GANs: firstly, incorporating Wasserstein GANs (WGANs) to prevent mode collapse, and secondly, conditioning these WGANs to secure accurate labels for the generated data. It is demonstrated that a basic FFNN, augmented with data yielded by a CWGAN, substantially improves its efficiency in scenarios where data is scarce, showing the inherent importance of CWGAN data augmentation beyond the simple expansion of datasets. Moreover, we present that CWGANs can function as a superior surrogate model when data is limited. Overall, this work aims at contributing to the research area of generative AI algorithms' applicability
beyond the conventional field of image generation.~We believe that our findings contribute to the understanding 
and applicability of generative AI algorithms in data-constrained contexts and could stimulate further 
research work in the application of generative AI in a variety of scientific scenarios.

\section*{Acknowledgements}
J.J.G.E.\ was supported by the Spanish Ministry of Science and Innovation through a FPU grant (FPU19/05281).
J.C.C.\ acknowledges funding from the Spanish Ministry of Science and Innovation (PID2020-114880GB-I00). 
J.B.A.\ acknowledges financial support from Ministerio de Ciencia, Innovaci\'on y Universidades (RTI2018-098452-B-I00).

\section*{References}
\bibliography{bibliography}

\end{document}